\documentclass[a4paper,10pt]{article}
\usepackage[margin=3cm]{geometry}
\usepackage{graphicx}
\usepackage{dsfont,amsmath,amssymb,color,comment}
\usepackage{graphics}
\usepackage{epsfig}
\usepackage{placeins}
\usepackage[comma,authoryear]{natbib}
\providecommand{\e}[1]{\ensuremath{\times 10^{#1}}}
\title{Tick size reduction and price clustering in a FX order book}
\author{Mehdi~Lallouache\thanks{Corresponding author. Email: mehdi.lallouache@ecp.fr} $^\textrm{a}$ and Fr\'{e}d\'{e}ric~Abergel$^\textrm{a}$}
\begin{document}
\date{\today}
\maketitle

\small
\begin{center}
$^\textrm{a}$~\emph{Chaire de Finance Quantitative, Laboratoire de Math\'{e}matiques Appliqu\'{e}es aux Syst\`{e}mes\\Ecole Centrale Paris, Ch\^{a}tenay-Malabry, 92290, France}

\end{center}
\normalsize

\vspace{1.0cm}
%%%%%%%%%%%%%%%%%%%%%%%%%%%%%%%%%%%%%%%%%%%%%%%%%%%%%%%%%%%%%%%%%%%%%%%%

\begin{abstract}
We investigate the statistical properties of the EBS order book for the EUR/USD and USD/JPY currency pairs and the impact of a ten-fold tick size reduction on its dynamics. A large fraction of limit orders are still placed right at or halfway between the old allowed prices. This generates price barriers where the best quotes lie for much of the time, which causes the emergence of distinct peaks in the average shape of the book at round distances. Furthermore, we argue that this clustering is mainly due to manual traders who remained set to the old price resolution. Automatic traders easily take price priority by submitting limit orders one tick ahead of clusters, as shown by the prominence of buy (sell) limit orders posted with rightmost digit one (nine).
\end{abstract}

\newpage

\section{Introduction}

The foreign exchange (FX) market, being the largest financial market in the world, affecting output, employment and inflation, rightly draws a lot of attention from academics. Since the 1990s, researchers have been able to access large datasets with intra-day resolutions. Starting with \citet{Wasserfallen1985} and \citet{Muller1990}, this wealth of data led to the emergence of a growing body of empirical studies on high-frequency FX rates. A set of stylized facts, reviewed for the first time by \citet{Guillaume1997} is now firmly established. The most important ones are the fat-tailed distribution of returns, the absence of linear autocorrelation of returns  (except on short time scales)  and the volatility clustering phenomenon. These properties are common to a wide range of assets: equities, commodities, bonds, etc. More recent and comprehensive results can be found in \citep{Dacorogna2001,Cont2001}.

The lack of readily available data on FX order books explains the relatively small number of empirical investigations on FX microstructure (beyond level I). Notable exceptions are \citet{Lo2008,Lo2010} and \citet{Kozhan2012}. These studies use data from Reuters, while our provider is EBS (Electronic Broking Service). These are the two largest FX electronic communication networks nowadays. Typically, dealing bank traders (and also hedge funds via prime brokerage) use EBS to conduct high volume transactions in order to liquidate unwanted accumulated inventory. For a detailed description of the foreign exchange market structure, see \citep{King2012}. Electronic interdealer trading represents around one third of spot FX trades \citep{BIS}. This market can be seen as the heart of the global FX market; therefore, it is relevant for FX order book studies. The EBS market with high-frequency data was already considered by several authors. \citet{Berger2008} investigate the relationship between order flow and exchange rates. \citet{Berger2009} analyze the factors driving the volatility persistence. Interestingly, they have shown that variations in market sensitivity to information play at least as large a role as do variations in the flow of information reaching the market through the trading process. \citet{Hashimoto2010} show that a "run"\footnote{Continuous increases or decreases in deal prices for the past several ticks.} has some predictive power on the direction of the next price move. Finally, the mechanisms behind FX rates tail events are studied by \citet{Osler2011}, who found that price-contingent trading may be a major source of extreme returns.

Here, we analyze data about two currency pairs: EUR/USD and USD/JPY that contains more quotes on each side of the book than the aforementioned studies. Crucially, they cover a period during which a major change of price resolution occurred. Indeed, in March 2011, EBS decided to reduce the tick size by a factor ten. More details about the data are provided in section \ref{part:Data}.

In this paper,  we want to take advantage of these features to analyze the order book's most important properties and see how they are affected by the change in tick size. \citet{Goldstein2000} analyzed similar tick size reduction in the equity context. The average shape of the book is deeply modified with the appearance of peaks at round distances and the spread distribution became bi-modal in the EUR/USD case.  The very high frequency sampling of our data (which makes them almost tick-by-tick in the EBS case) allows us to devise a method to infer the stream of orders (limit orders and cancellations) from the deal and quotes data. This reveals strange patterns in the order placement and volume. We show that these facts stem from the emergence of a strong price clustering, i.e. a tendency for prices to congregate around some specific values, after the tick size reduction. Surprisingly in such a liquid and mature market, the clustering is very strong and stable in time. \citet{Goodhart1991,Sopranzetti2002} and \citet{Mitchell2006} also noticed clustering in spot FX rates but their studies concern the pre-euro era and use low-frequency indicative quotes\footnote{Indicative quotes are non-binding quotes that have been posted by individual banks to the electronic data networks for informative purpose.}. \citet{Osler2003} reports clustering in stop-loss and take-profits orders placed at a large dealing bank (National Westminster Bank). Our study shows that the phenomenon is pervasive in the EBS market. Clustering affects transaction prices and order prices. It creates an accumulation of volume at round numbers turning them into price barriers, thus presumably hindering markets' ability to process information efficiently. Finally, we show that the clustering is mainly due to manual traders who do not use the new price resolution. Automatic traders (computer algorithms) take advantage of this behavior by submitting buy (sell) limit orders just above (below) prices ending with a zero or a five, thus easily taking price priority over large clusters.

The plan for the remainder of this paper is as follows. Section 2 describes the data we use. Section 3 reports basic results on the order book, presents the orders reconstruction procedure and describes the characteristics of the resulting orders. Section 4 documents the existence of price clustering in EBS and provides evidence on its origin. Section 5 concludes.

\section{Data}
\label{part:Data}

Most of spot interdealer trading occurs on two competing platforms: EBS Spot and Reuters D-3000. Due to network externalities, liquidity naturally gravitated to just one platform for each currency. EBS has long dominated interdealer trading for the EUR, JPY, and CHF, while Reuters dominates the GBP, AUD, CAD, and the Scandinavian currencies. In this paper, we study two major currency pairs for which EBS is the leader: EUR/USD and USD/JPY. The following periods of historical data were bought from EBS: from August 1, 2010 to July 31, 2011 and from January 1, 2012 to March 31, 2012. The dataset contains a \emph{Quote Record}   and a \emph{Deal Record} on a $0.1 s$ time-slice basis. The Quote Record is a snapshot of the ten best levels of the book at the end of a time-slice (if a price or a volume in the book changed within the time-slice). The Deal Record lists the highest buying deal price and the lowest selling deal price (with the dealt volumes) during the time-slice. For the second time-period, we also know the total signed volume of trades in a time-slice. This is so far the best available data from EBS in terms of frequency (almost tick by tick) and depth (10 levels). In March 2011,  EBS decided to set in a tick \emph{decimalization}. The EUR/USD tick size changed from $10^{-4}$ to $10^{-5}$ and  the USD/JPY one from $10^{-2}$ to $10^{-3}$. In FX terminology, the market went from \textit{pip-pricing} to \textit{decimal pricing}.

\section{The limit order book}
\label{part:book}

Most of today's financial markets use a limit order book mechanism to facilitate trade. Thanks to the computerization of markets, researchers can access extensive data on order books allowing them to put the market under the microscope. The price dynamics emerges as a complex interplay between the order book and the order flow. Starting with \citet{Biais1995}, researchers from different fields are getting new insights to understand this complexity \citep{Maslov2001,Bouchaud2002,Zovko2002}. A recent review can be found in Ref. \citep{Chakraborti2011a}. In this section, we use the EBS data to revisit the basic order book results. Our motivation is two-fold. First, as we mentioned in the introduction, few studies deal with the FX order book. Second, a tick decimalization occurred in March 2011, which is part of our dataset; therefore, we can study the reaction of the market to this change. Note that intraday seasonalities are a well-known feature of FX data \citep{Dacorogna2001,Ito2006}, therefore in the following, we will use London opening hours only (8 a.m. - 6 p.m.).

\subsection{Average shape of the order book}

\begin{figure}
\begin{center}
\includegraphics[width=0.4\columnwidth]{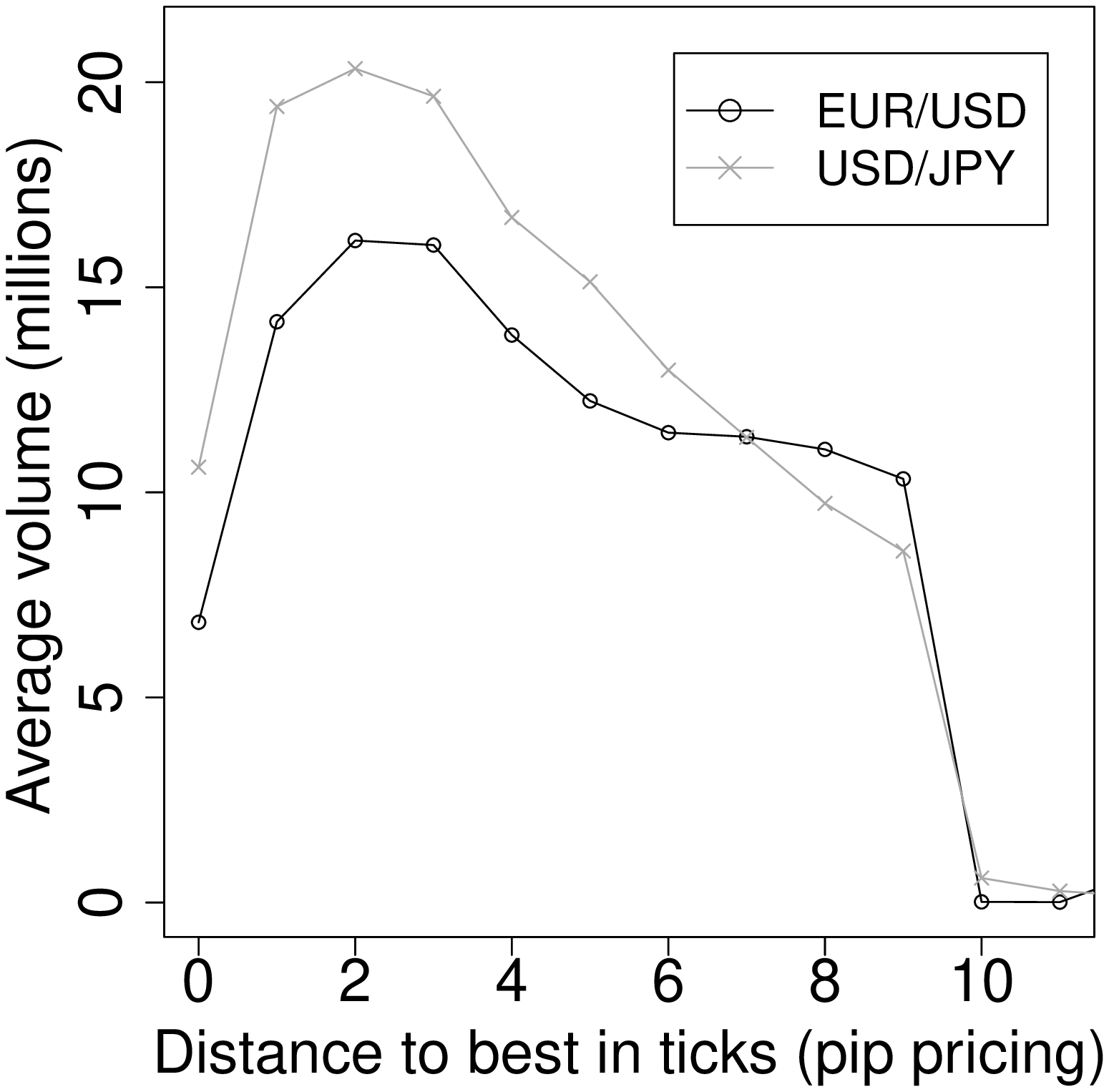}
\includegraphics[width=0.4\columnwidth]{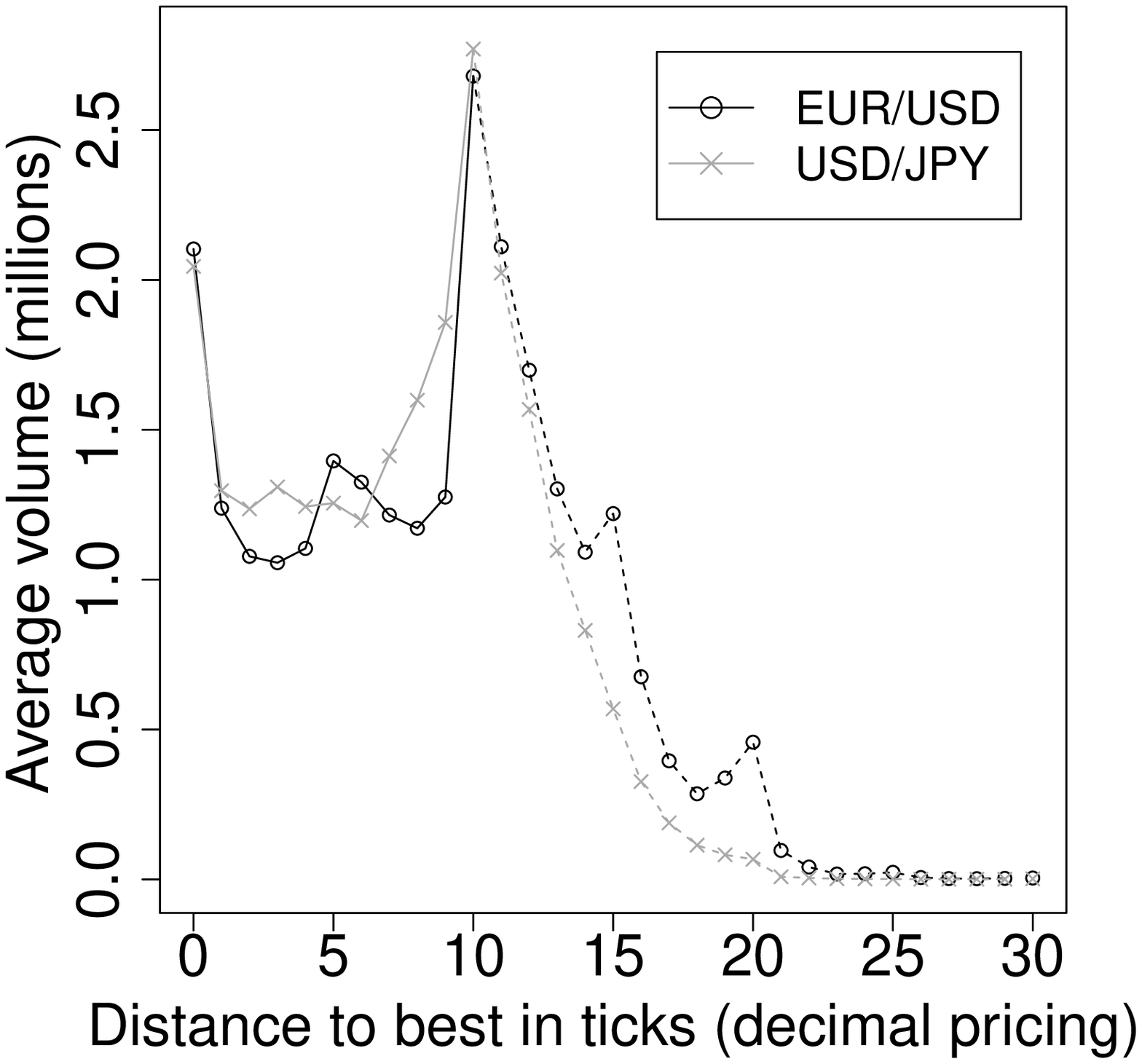}
\end{center}
\caption{Average shape of the book ask-side. Similar shape for the bid side. (Left) Before decimalization: February 2011. (Right)  After decimalization: March 2012. The shape is exact up to 10 ticks}
\label{figure:shape}
\end{figure}

A usual way to represent the shape of the book is to compute the (physical) time-averaged volume in the order book as a function of the distance from the current bid (or ask). In figure \ref{figure:shape} we plot the shape before and after the decimalization. As already highlighted by other studies (see, e.g., Ref. \citep{Bouchaud2002} for the equity market) the maximum is not located at the best quote. Before the decimalization, we can see the well-known hump-shaped curve with a maximum at two pips. The volume after ten ticks seems very small because we only have access to the ten first levels. After decimalization, the shape is unusual. The volume decreases after the best quotes and then increases to reach a maximum at $10$ (which corresponds to one pip). For EUR/USD we can also see small peaks at $5$,$15$ and $20$ ticks. We investigate these peculiarities in section \ref{part:clustering}. The results are similar for the bid side and for others months in the dataset.

Another interesting quantity related to the shape of the book is the average gap, i.e. the price distance between two levels. We plot this quantity in figure \ref{figure:gs} after decimalization. Before decimalization, the gap is almost always equal to one-tick. After decimalization, it decreases with the level. The results are similar for the ask side and do not change in time.

\subsection{Spread}
\label{part:spread}

\begin{figure}
\begin{center}
\includegraphics[width=0.4\columnwidth]{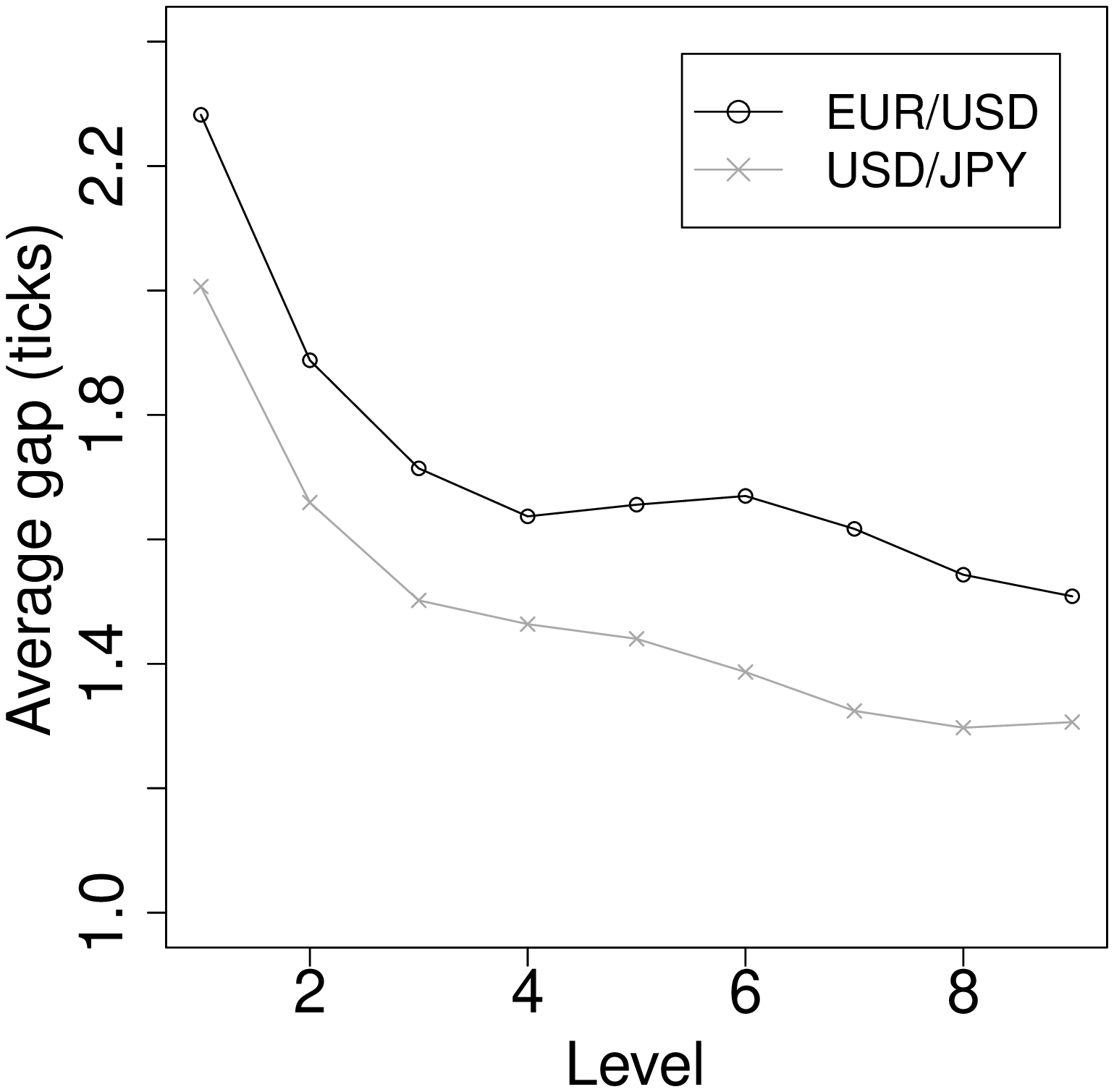}
\includegraphics[width=0.4\columnwidth]{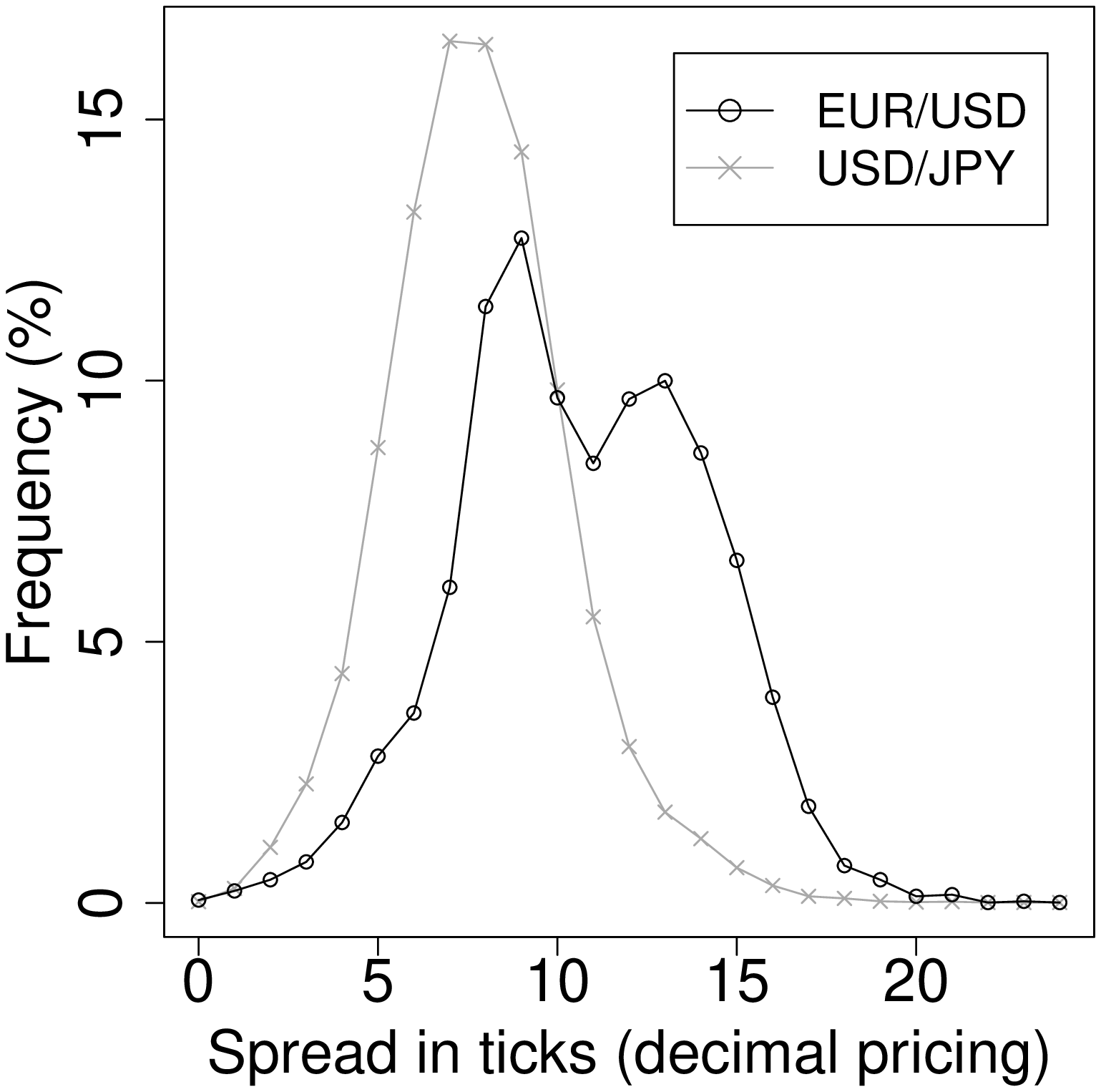}
\end{center}
\caption{ (Left) Average gap in the book bid-side. Similar results for the ask side. (Right) Spread distribution. Data from March 2012.}
\label{figure:gs}
\end{figure}

One of the most important quantity for traders is the difference between best ask and best bid, called the \emph{spread}, because it measures the cost of making a transaction immediately through a market order. Before decimalization, the spread was equal to one tick $65 \%$ of the time and to two ticks otherwise. We want to know the extent of the impact of the decimalization on the spread. For this purpose, we compute the spread distribution using one-month data sampled every second from 8 a.m. to 6 p.m. The results are presented in figure \ref{figure:gs}. The spread can now take many values around the previous typical spread: $10$. For the EUR/USD the distribution presents a bimodality. The first mode is at $9$ and the second is at $13$. The second mode may change slightly depending on the considered time period, but we always have the first mode at $9$ and the second one greater than $10$. The results are similar for different sampling frequencies. The USD/JPY distribution seems more "natural" and the USD/JPY spread is smaller than the EUR/USD one. We explain the bimodality in section \ref{part:spread_clust}.

The two previous subsections illustrate the change from a large tick asset to a small tick asset. The spread and the gaps switched from one tick to a few ticks and the volume at each level was reduced.  In other terms, a very dense book became sparse (in ticks, not in absolute prices).

\subsection{Order reconstruction}

EBS does not publish data on the submission of limit orders and cancellations. Nevertheless, we can infer order submissions from deal and quote messages in our high-frequency data. Perfect reconstruction is impossible: the 10-levels limitation and the $0.1 s$ time-slicing imply loss of information. Nevertheless, in most cases nothing happens within a time-slice. For example, between 2 p.m. and 4 p.m. (two most active hours of the day) nothing changes in the book in half of the time-slices. Moreover market participants also face this 0.1 s time-slicing so they cannot act at a much higher frequency. We therefore  capture most of the order book changes. To build the order flow, we first split the data according to their  side (bid or ask) and then we go sequentially through the quotes and compare two subsequent snapshots. When there is no transaction between two subsequent snapshots, all the changes between the snapshots are easily explained by limit orders and cancellations. When there is a transaction within a time-slice, we face two cases. Case 1 (around 75 per cent of the cases): the total traded volume is equal to the reported trade\footnote{By reported trade, we mean the lowest selling deal or highest buying deal depending on the considered side, see section \ref{part:Data}} volume. In this case, we know that there is a unique trade in that time-slice (for the considered side), then we match the deal volume with the corresponding volume decrease between the two corresponding subsequent snapshots. The rest of the liquidity changes are explained by limit orders and cancellation. Case 2: the total traded volume is greater than the reported trade volume. We proceed like the previous case for the reported trade and we randomly distribute the remaining dealt volume among the available prices (from best price to the reported trade price). Again the remaining  liquidity changes are attributed to limit orders and cancellations. Some orders of magnitude obtained with this procedure are given in Table \ref{table:OM} along with the number of deals for comparison.

\begin{table}
\begin{center}
\begin{tabular}{lrrrrrr}
\hline
\hline
\multicolumn{1}{l}{}&\multicolumn{1}{c}{Limit orders}&\multicolumn{1}{c}{Cancellations}&\multicolumn{1}{c}{Trades}\tabularnewline
\hline
EUR/USD&$220$&$200$&$25$\tabularnewline
USD/JPY&$75$&$70$&$7$\tabularnewline
\hline
\end{tabular}
\end{center}
\caption{Average number of events per day (in thousands).}
\label{table:OM}
\end{table}

\subsection{Order characteristics}

The total traded volume is only available for 2012 data so we cannot look into the order characteristics before the decimalization.

\subsubsection{Volume}

\begin{figure}
\begin{center}
\includegraphics[width=0.5\textwidth,angle=-90]{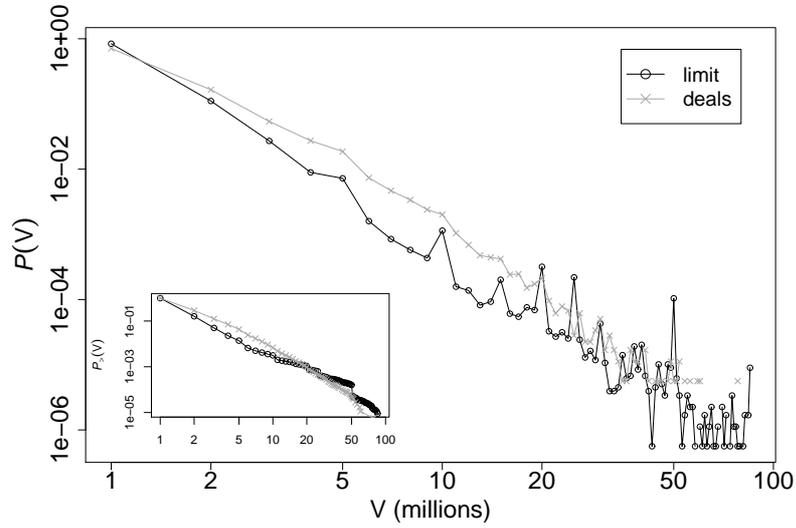}
\end{center}
\caption{EUR/USD orders volume distribution in log-log scale. Similar results for USD/JPY. Sample period: march 2012. Inset: Cumulative distribution.}
\label{figure:size}
\end{figure}

In the EBS market, order size must be a multiple of $1$ million (of the base currency). Figure \ref{figure:size} shows the distribution of EUR/USD order volumes (data from March 2012). The majority (around $80 \%$) of the order sizes is at the minimal value (1 million euros). We observe a strong representation of limit orders and cancellations around 5, 10, 15, 20, 25, etc. Although weaker, the effect is also present for deals volumes. A similar round numbers preference has been found for trade sizes in the Chinese stock market \citep{Mu2009}. These peaks are explained in section \ref{part:twotypes}. The results are stable through  time and are similar for USD/JPY.

\subsubsection{Placement}

\begin{figure}
\begin{center}
\includegraphics[width=0.4\columnwidth]{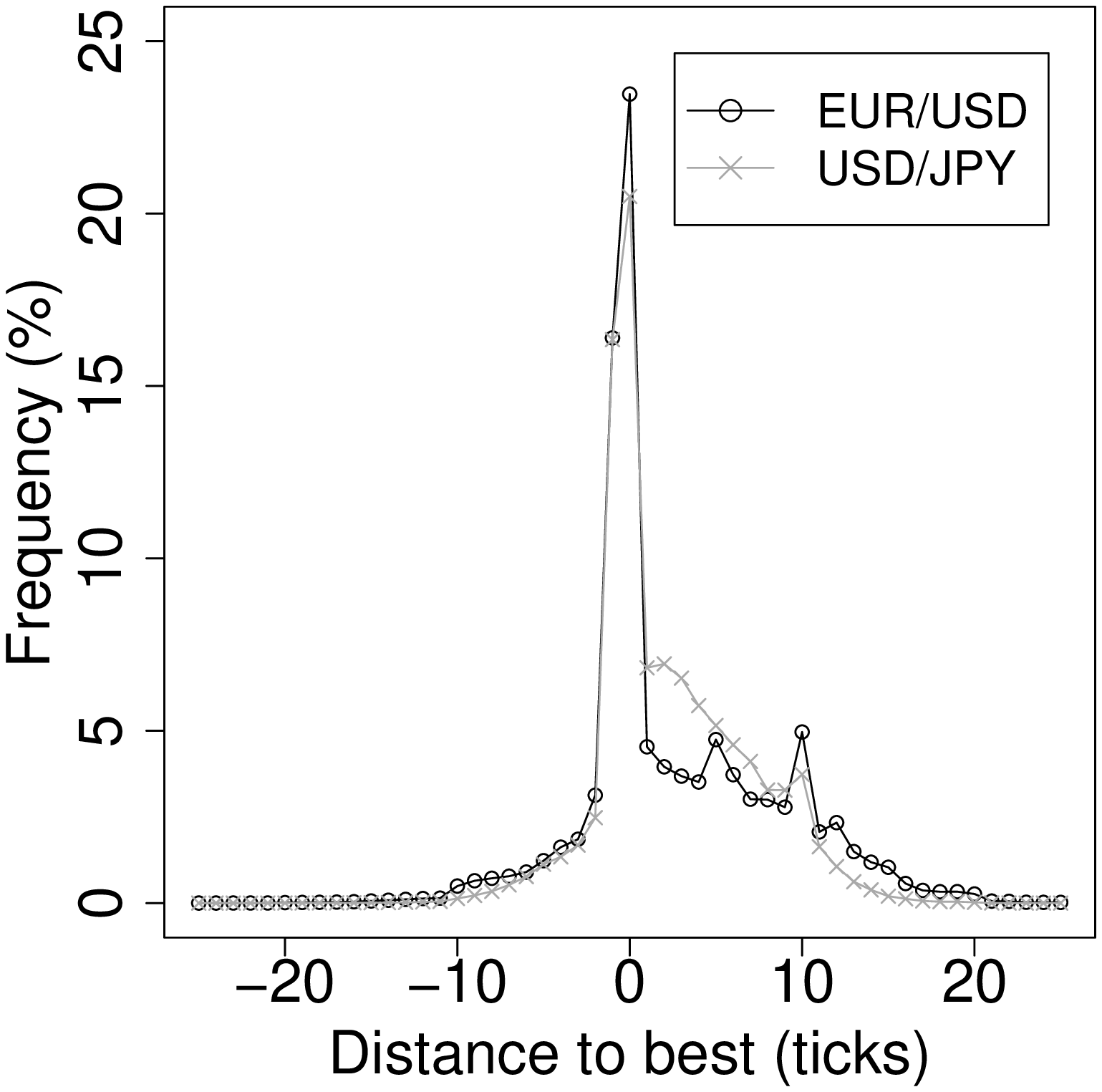}
\includegraphics[width=0.4\columnwidth]{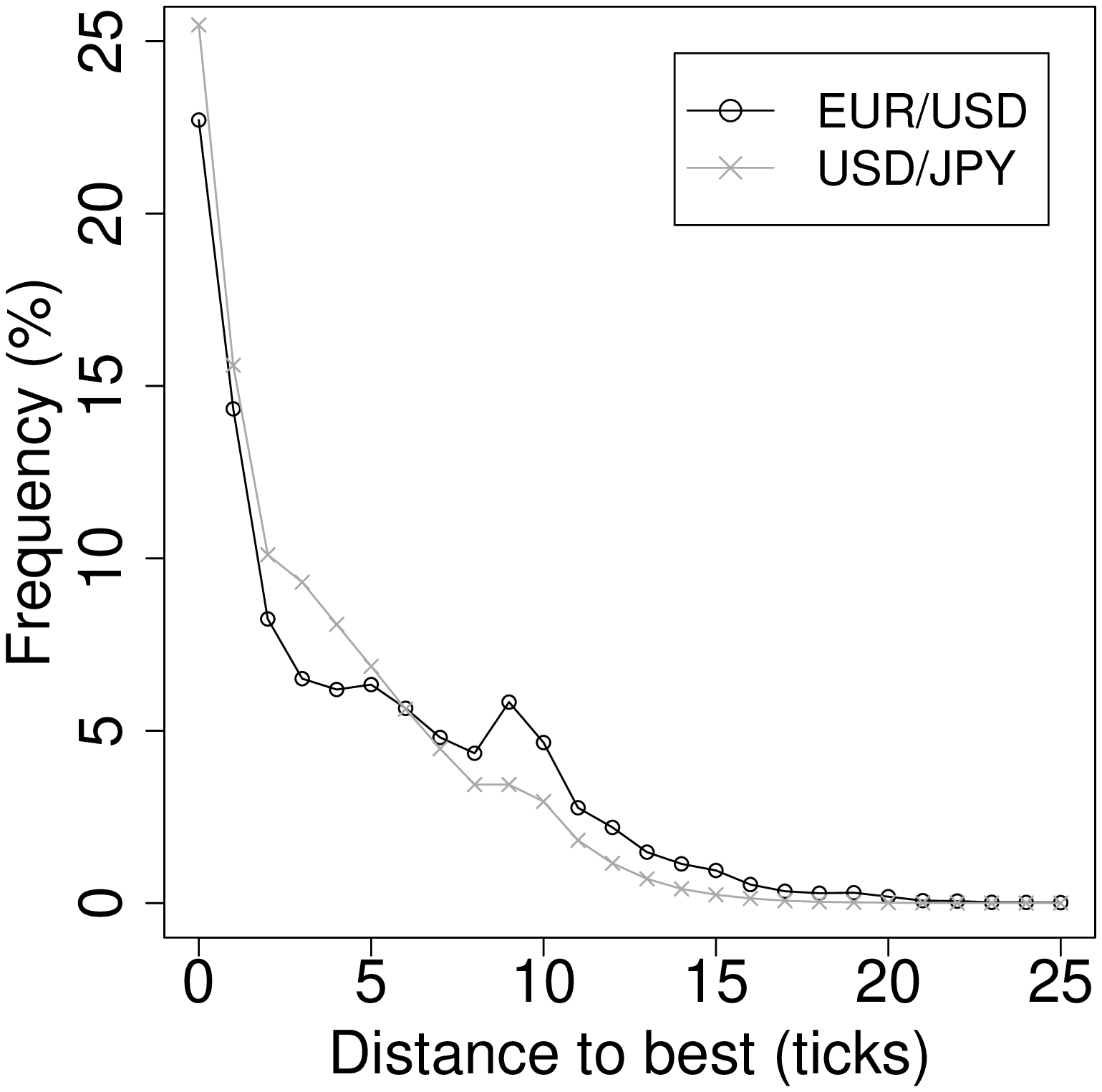}
\end{center}
\caption{Placement of orders using the same best quote reference for February 2012. (Left) Limit orders. (Right) Cancellations. Ask side and bid side are similar.}
\label{figure:placement}
\end{figure}

Let $\delta$ be the distance between the current price and an incoming limit order price. More precisely: ${\delta = b_0(t-) - b(t)}$ (resp. ${a(t) - a_0(t-)}$) if a bid (resp. ask) order arrives at price $b(t)$ (resp. $a(t)$), where $b_0(t-)$ (resp.$a_0(t-)$) is the best bid (resp. ask) before the arrival of this order. A classic interesting question concerns the distribution of $\delta$. Results for EBS are plotted in figure \ref{figure:placement}.

These graphs being computed with incomplete data (ten best limits), we do not observe a placement as broad as in Ref. \citep{Bouchaud2002}. The empirical distribution is asymmetric: the left side is less broad than the right side. Since the left side represents limit orders submitted \emph{inside} the spread, this is expected. The distribution has a maximum located at the current best price ($\delta = 0$) and a high value at $-1$, which is the smallest price improvement. In the EUR/USD case, we notice two clear peaks at $5$ and $10$, which correspond to a half-pip and a pip distances. The origin of these peaks is discussed in section \ref{part:clustering}.

\subsubsection{Arrival times}

\begin{figure}
\begin{center}
\includegraphics[scale=0.4]{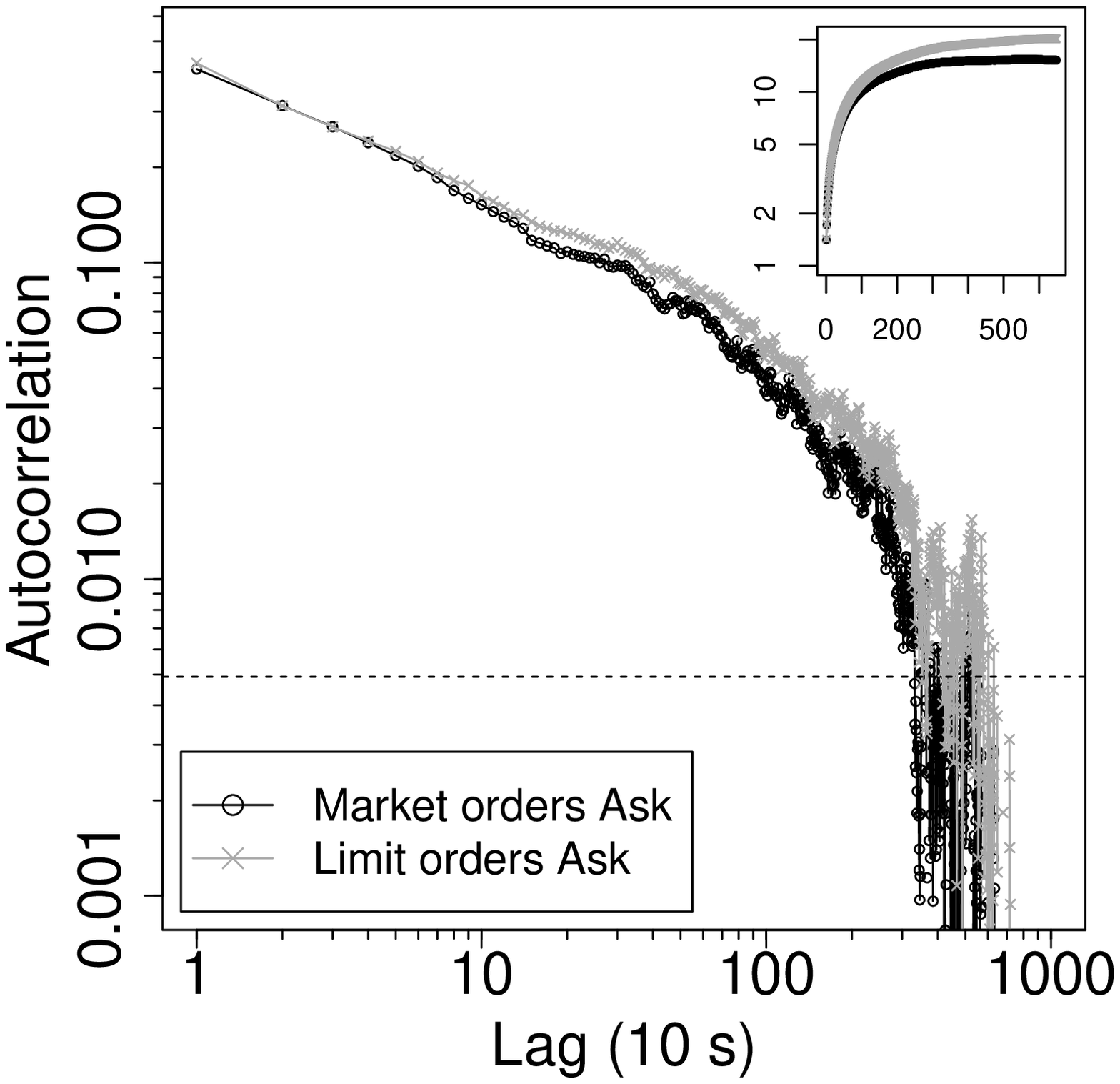}
\includegraphics[scale=0.4]{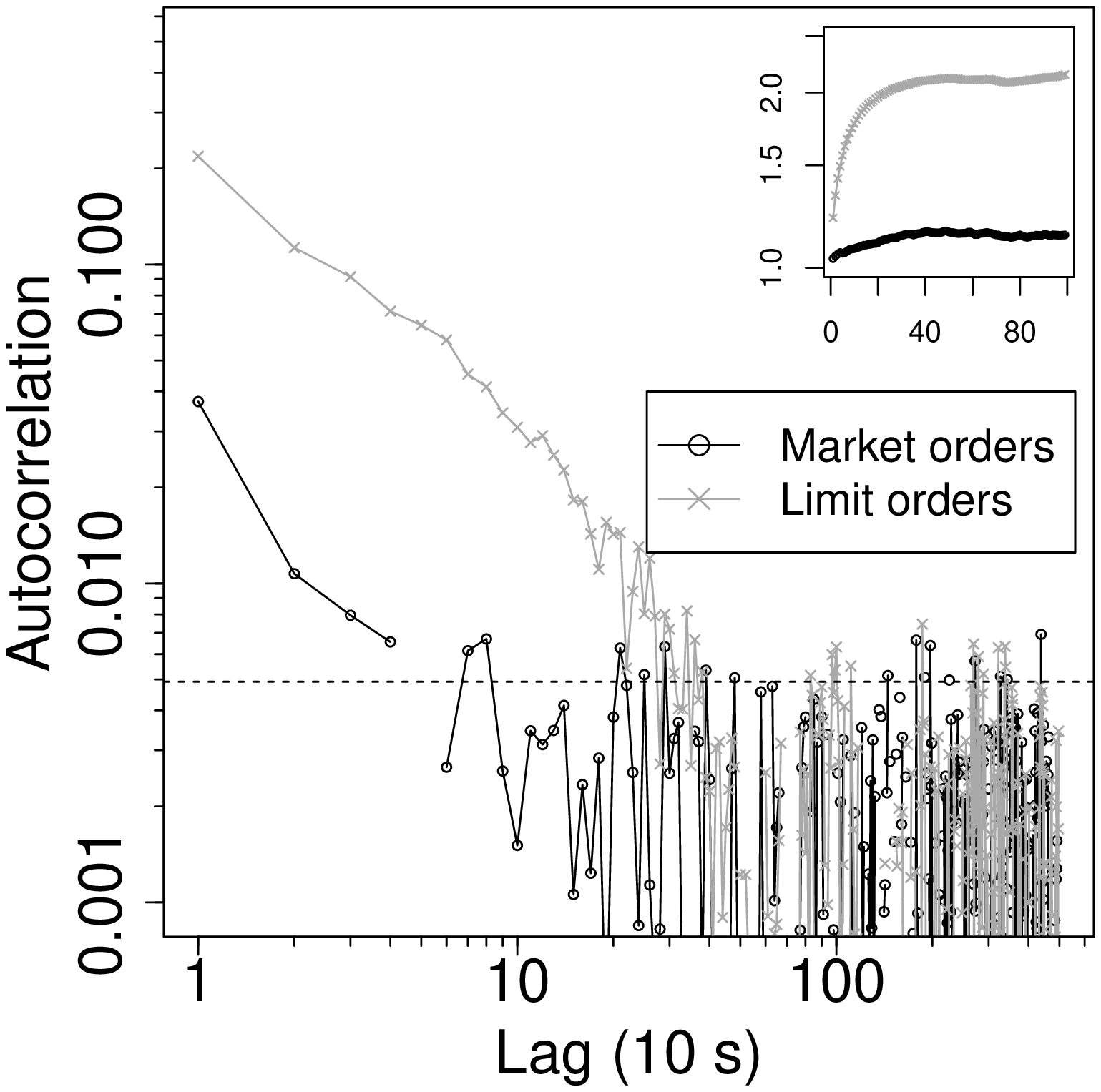}
\end{center}
\caption{ (Left) Autocorrelation function of the number of orders in a $10$ s window. (Right) Autocorrelation function of the mean sign in a $10$ s window. Cancellations and limit orders are similar. Insets: Cumulative sum in semi-logarithmic scale. Sampling period: January and February 2012.}
\label{figure:acf_orders}
\end{figure}

It is now clearly established that the Poisson hypothesis for the arrival times of orders is not empirically verified  (see \citep{Chakraborti2011a} and references therein). The data resolution prevents us to study directly the inter-arrival times but we can compute the number of events in a $10$ s window\footnote{We chose the window length as a trade-off between the number of windows and the number of empty windows .}. Let us consider 6 types of events: limit orders, market orders and cancellations on each side of the book and investigate clustering and inter-dependence phenomena among them. We restrict ourselves to linear dependencies with the autocorrelations and cross-correlations for the time-series of the number of events. Figure \ref{figure:acf_orders} (left) shows that the autocorrelation is statistically different from $0$ at several lags. The cumulative sum of the autocorrelation coefficients, which saturates for large lags, shows that the generating process does not have long-memory. Table \ref{table:events} is the correlation matrix for the six time-series. Significant correlations (as high as $0.84$) are present, demonstrating the inter-dependence  between order arrival processes. The independent Poisson processes hypothesis is clearly rejected. As suggested by recent studies, Hawkes processes are better candidates for orders time arrival modeling.

\begin{table}
\begin{center}
\begin{tabular}{lrrrrrr}
\hline\hline
\multicolumn{1}{l}{}&\multicolumn{1}{c}{Limit Ask}&\multicolumn{1}{c}{Cancel Ask}&\multicolumn{1}{c}{Market Ask}&\multicolumn{1}{c}{Limit Bid}&\multicolumn{1}{c}{Cancel Bid}&\multicolumn{1}{c}{Market Bid}\tabularnewline
\hline
Limit Ask&$1.000$&$0.837$&$0.561$&$0.563$&$0.750$&$0.716$\tabularnewline
Cancel Ask&$0.837$&$1.000$&$0.602$&$0.751$&$0.716$&$0.571$\tabularnewline
Market Ask&$0.561$&$0.602$&$1.000$&$0.726$&$0.585$&$0.644$\tabularnewline
Limit Bid&$0.563$&$0.751$&$0.726$&$1.000$&$0.841$&$0.566$\tabularnewline
Cancel Bid&$0.750$&$0.716$&$0.585$&$0.841$&$1.000$&$0.600$\tabularnewline
Market Bid&$0.716$&$0.571$&$0.644$&$0.566$&$0.600$&$1.000$\tabularnewline
\hline
\end{tabular}
\end{center}
\caption{Correlation matrix of the number of events time-series for EUR/USD. Similar values for USD/JPY.}
\label{table:events}
\end{table}

\subsubsection{Signs memory}
\citet{Lillo2003} demonstrated that the signs of orders in the London Stock Exchange obey a long-memory process. In a similar fashion, we compute the mean sign ($+1$ for buy orders, $-1$ for sell orders) in a $10$ s window for limit orders, cancellations and market orders and look at the autocorrelation function of these time-series. The results are plotted in figure \ref{figure:acf_orders} (right). Contrary to the equity market, the market order autocorrelation function is rapidly decaying and can be considered null after about $2$ minutes. This result was communicated to us by \citet{Curato2014}. The autocorrelation function for limit orders (and for cancellations) displays a slower decay and becomes statistically zero after approximately $5$ minutes. 

\section{Price clustering}
\label{part:clustering}
Using previously gathered empirical facts, especially the average shape of the book and the limit order placement in section \ref{part:book}, we can anticipate a strong price clustering due to the tick decimalization. Price clustering is the tendency for prices to center around certain values. Many empirical studies have revealed that investors do not fully use the price resolution allowed by the tick size. The first statistical investigation on this phenomenon was the work by \citet{Osborne1962} followed by \citet{Niederhoffer1965,Niederhoffer1966}. Since then, the focus has been mainly on equity markets \citep{CooneyJr.2003,Ahn2005,Cellier2007,Ikenberry2008,Onnela2009}, the relevant foreign exchange literature is given in the introduction. There are a number of proposed explanations regarding this clustering property. The price resolution hypothesis \citep{Ball1985} argues that the degree of clustering varies inversely with the information about the underlying value of the asset. If the value is well known, traders will use a finer price grid. The negotiation hypothesis \citep{Harris1991} posits that traders coordinate to restrict themselves to a smaller set of prices in order to reduce negotiations costs. The attraction hypothesis \citep{Goodhart1991} states that investors have a natural attraction towards round numbers. The collusion hypothesis \citep{Christie1994,Christie1994a} asserts that dealers avoid certain prices (odd-eights in the NASDAQ case) to maintain artificially wide spreads. In the EBS market, we show that the clustering is due to specific patterns in limit order placement. For liquid currency pairs, if the tick size is appropriate, no clustering should occur. As we are interested in clustering due to the decimalization, we are going to focus on the price last digit\footnote{We have checked that there is no clustering on the other digits. Of course the first digit seems clustered but it is just an intrinsic value of the exchange rate. It is very unlikely that EUR/USD rises above $2$ for example.}. A spurious bid/ask asymmetry may appear if one looks at the last digit directly on both sides.  For example, let us suppose that the best quotes are on prices whose last digit is 0, which we define as \textit{integer prices}. In this situation, if a limit order improves the best quote by one tick at the bid, its last digit is $1$, whereas if it is posted at the ask, its last digit is $9$. In our view, it is the same situation and should give the same "decimal part". In the following, we will use the term last digit on both sides, but when it concerns a price on the ask side, it will actually designate the distance (number of ticks) to the smallest integer price bigger than the price.

\subsection{Trade price}

\begin{figure}
\begin{center}
\includegraphics[width=0.4\columnwidth]{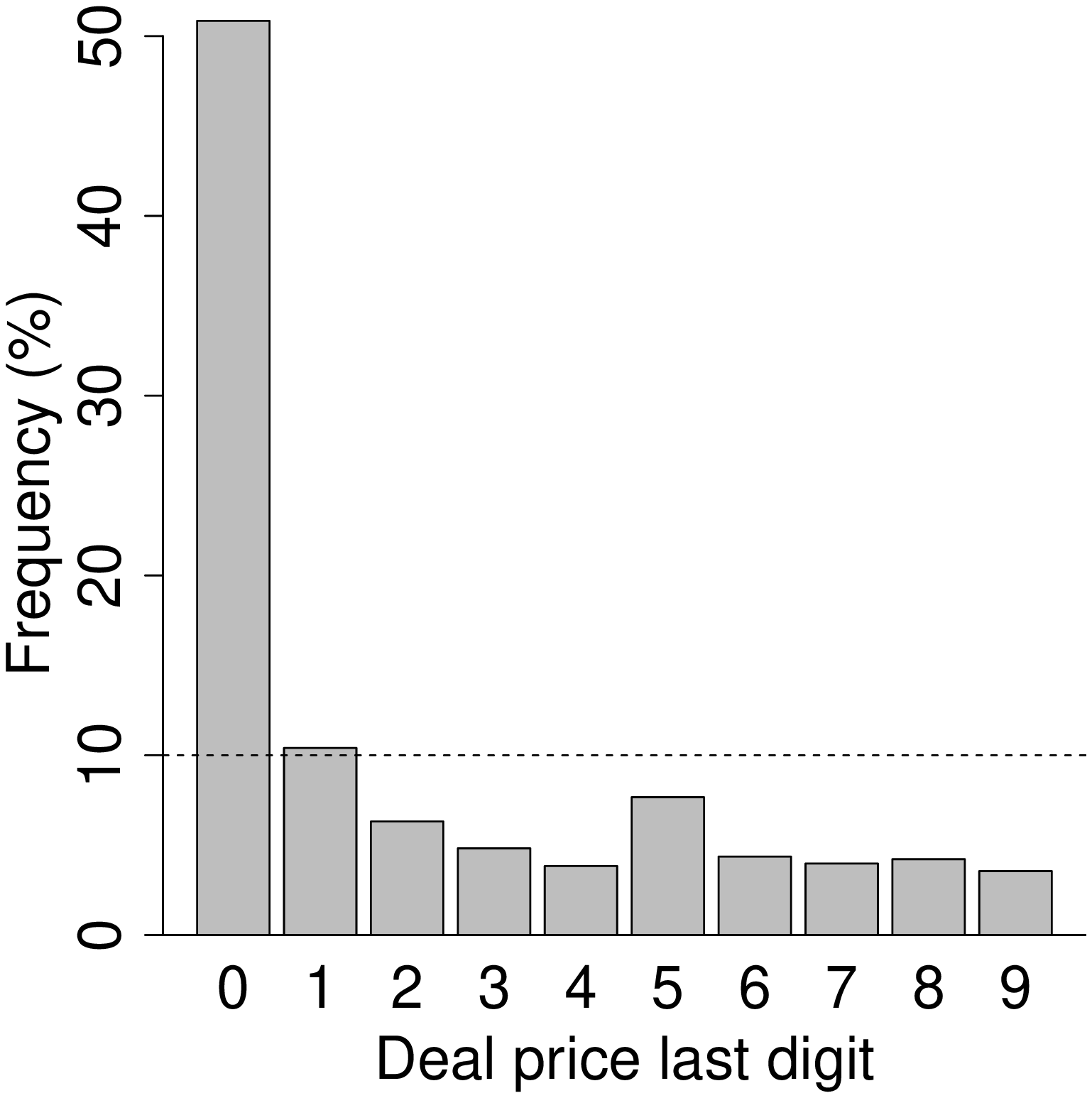}
\includegraphics[width=0.4\columnwidth]{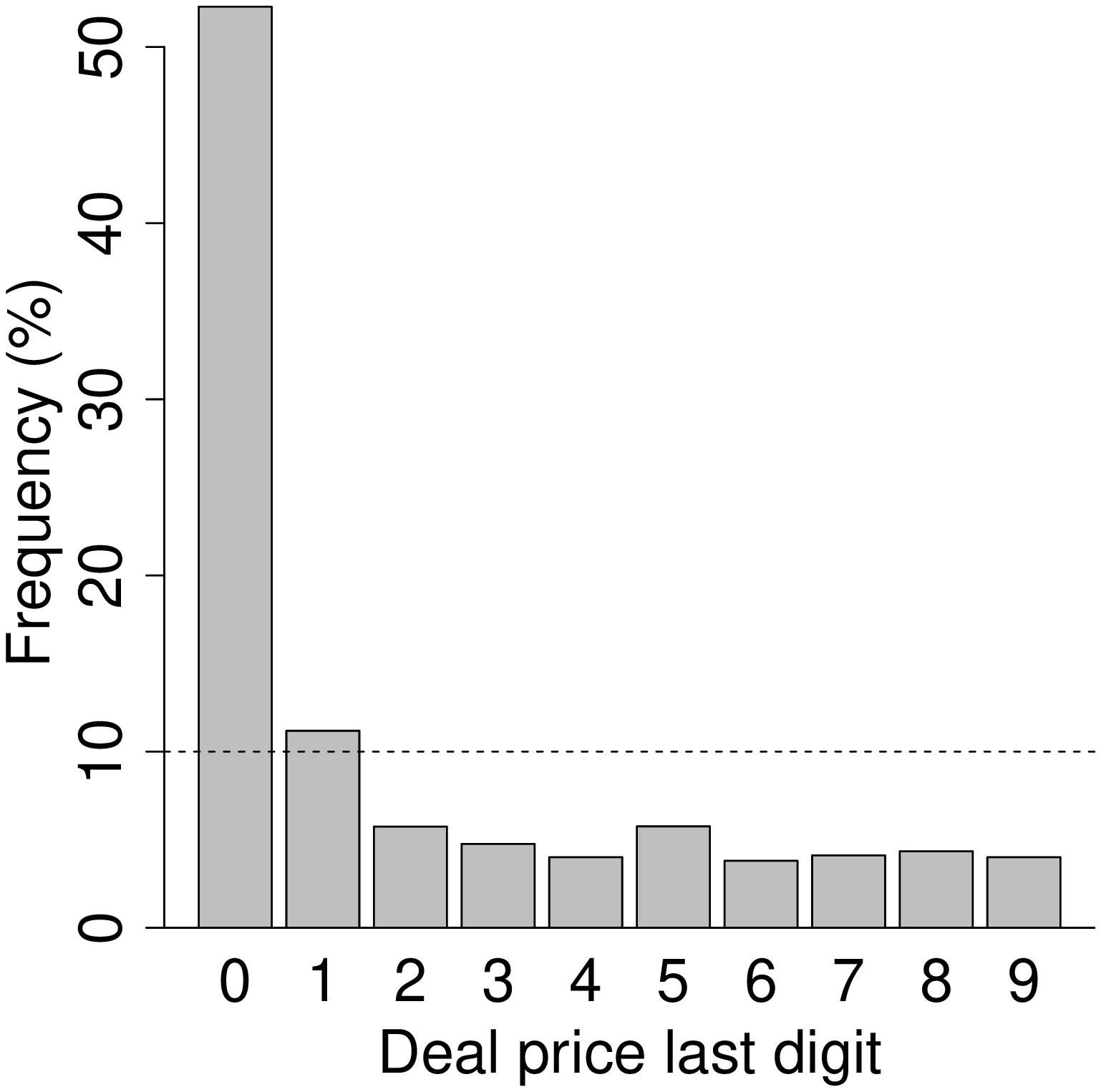}
\end{center}
\caption{Trade prices (ask side) last digit distribution for March 2012. The dashed line represents the theoretical frequency under the uniform hypothesis.  (Left) EUR/USD. (Right) USD/JPY. 
Around $50 \%$ of the trades occurs at integer prices. Same results for bid side.}
\label{figure:deal_clust}
\end{figure}

During periods of active trading it is natural to assume that realized trades should not cluster at certain prices. Under this assumption, the distribution of the last digit of price should be uniform. Figure \ref{figure:deal_clust} plots the frequencies $\hat{p_i}=\frac{n_i}{n}$, where $n_i$ is the number of trades with last digit $i \in \{0,1,...,9\}$ and $n$ is the total number of trades. The measurement uncertainties can be estimated by $1.96[\frac{\hat{p_i}(1-\hat{p_i})}{n}]^{\frac{1}{2}}$. They are all smaller than $0.0016$, so we can visually assess that the frequencies are very far from uniformity. To confirm this statement, we performed $\chi^2$ tests on different months and the uniform distribution hypothesis is always rejected at the $1\%$ level\footnote{This is true for the two following subsections.}. Deal prices with a $0$ as last digit (integer prices) represent about $50\%$ of the trades. In other words, the old tick is somehow still present. We also checked that the distribution was uniform before the decimalization.

\subsection{Limit orders}
\label{part:limit_clust}
\begin{figure}
\begin{center}
\includegraphics[width=0.4\columnwidth]{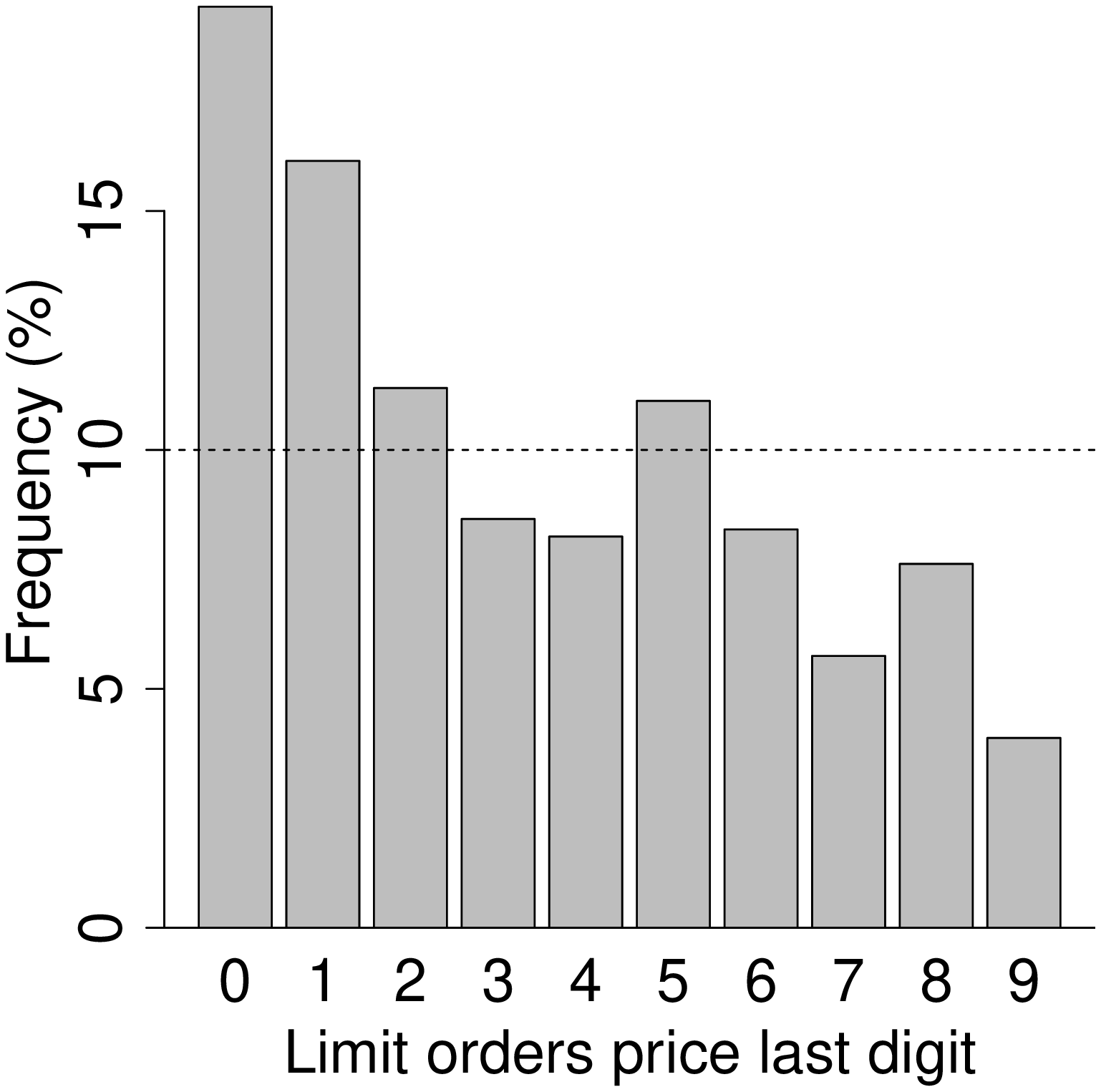}
\includegraphics[width=0.4\columnwidth]{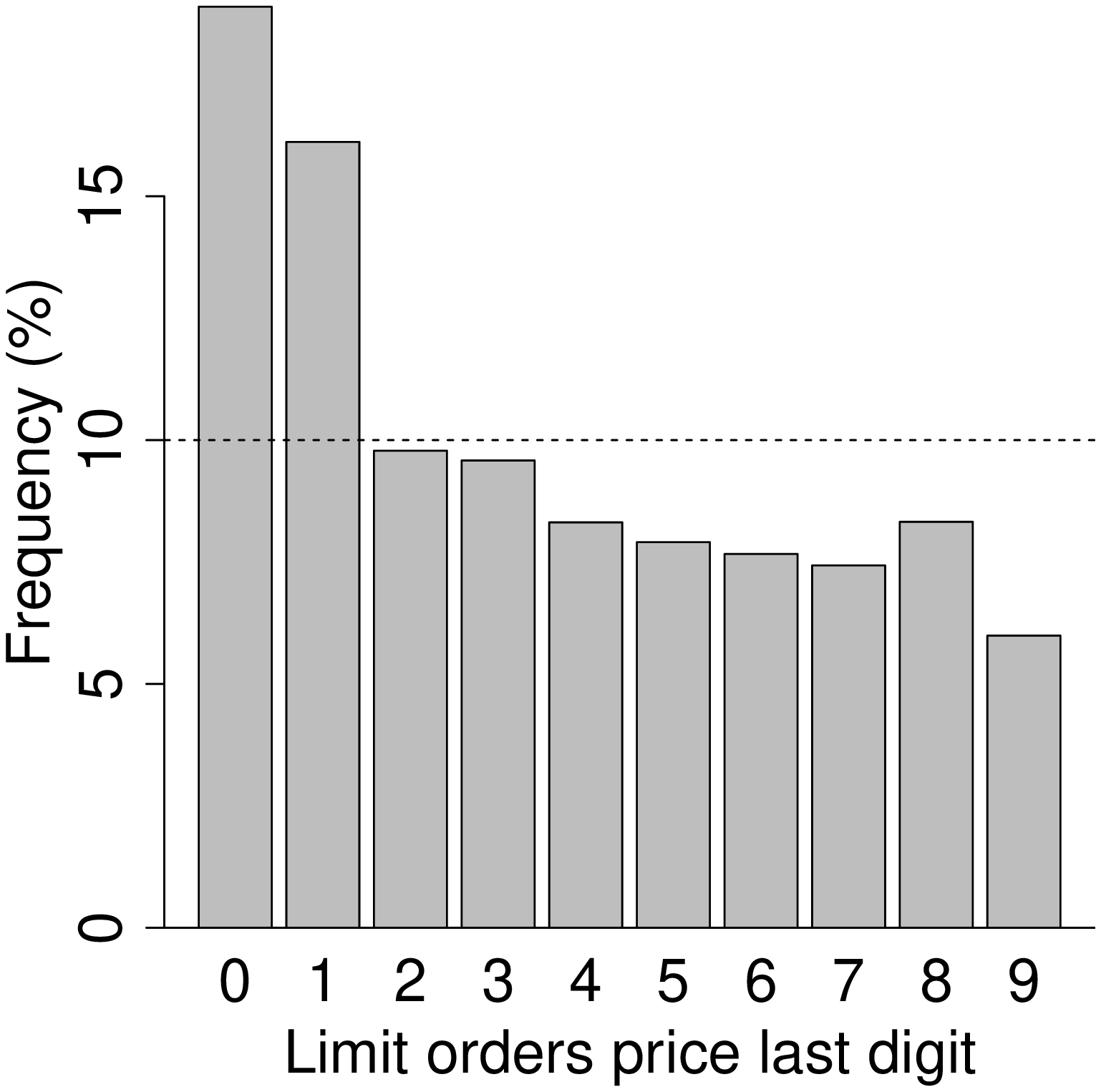}
\end{center}
\caption{Distribution of last digits in limit order prices in March 2012.  (Left) EUR/USD. (Right) USD/JPY. Around $20 \%$ of the orders are posted at integer prices.}
\label{figure:limit_clust}
\end{figure}

We now look at price clustering for limit orders. The last digit frequencies for limit order prices are plotted in figure \ref{figure:limit_clust}. The frequency uncertainties order of magnitude is $5\e{-4}$. Again, the fractions are not equally frequent. The clustering is less pronounced than in trade prices but it is still strong. Around $20 \%$ of the orders are posted at integer prices and a half-integer peak is present in the EUR/USD case. The next prominent last digit is $1$, this is certainly related to the strategic behavior of some traders, who anticipate clustering tendencies and step-ahead round prices to obtain priority.

The limit order relative price distribution peaks or the average shape peaks (both at $5$ and $10$, see figure  \ref{figure:placement} and \ref{figure:shape}) might rise the question: is there also a round distance preference? The answer is negative, these peaks come from the price clustering. We verified this by computing the aforementioned distributions conditionally on the best quote last digit. The peaks positions change in a way that favors round prices and not round distances. The depth accumulation at $5$ and $10$ in the average shape comes directly from price clustering.

\subsection{Best quote - Price Barriers}

\begin{figure}
\begin{center}
\includegraphics[width=0.4\columnwidth]{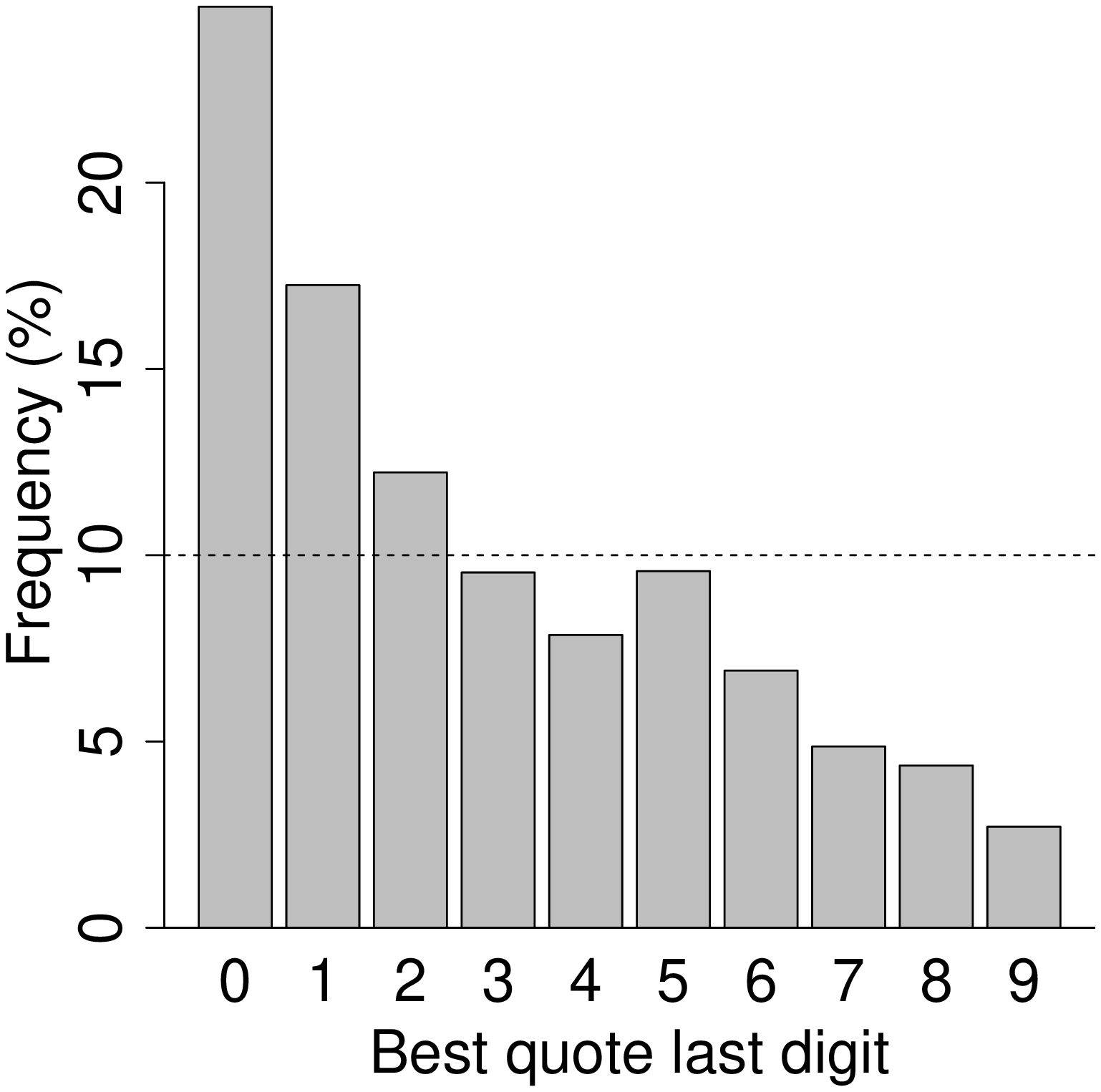}
\includegraphics[width=0.4\columnwidth]{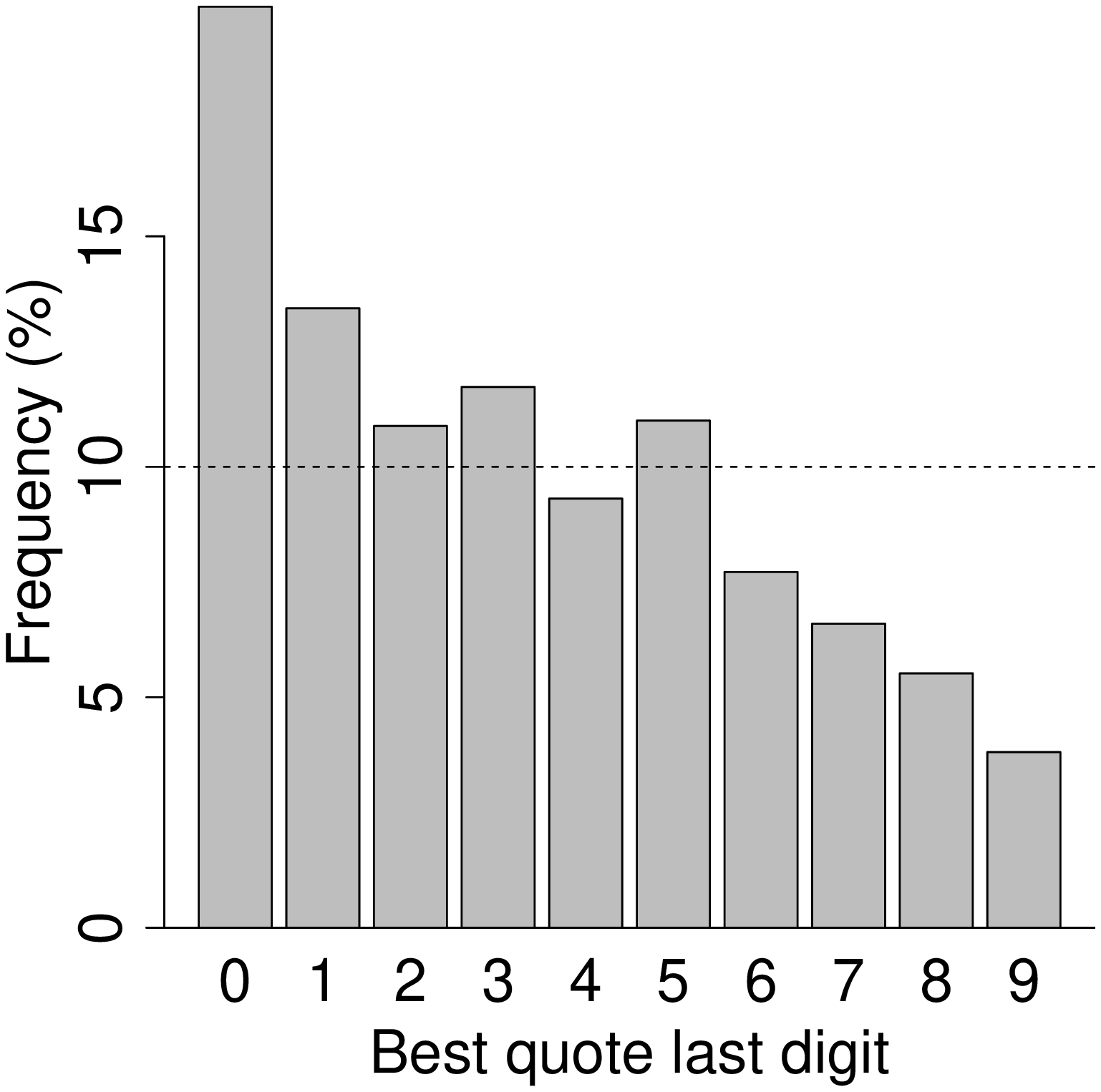}
\end{center}
\caption{Best quote last digit distribution for January 2012 ($1$ s sampling). (Left) EUR/USD. (Right) USD/JPY. The best bid (or ask) spend most of the time on integers.}
\label{figure:best_clust}
\end{figure}

Limit orders have most of the time a size of $1$ million (figure \ref{figure:size}), then a clustering in terms of number of orders is also a clustering in terms of volume. Consequently, price clustering generates depth accumulation at round prices, affecting the best quote dynamics. Since more volume will be necessary to push the price through integers and halves, round best quotes may constitute "price barriers". The fact that there are more transactions at round prices may offset the depth accumulation. We show that this is not the case by recording the best quote last digit every $0.1$ s and plotting its distribution in figure \ref{figure:best_clust}. A congestion effect is present, since more time is spent on round prices. This is an important point since it shows that a change in a microstructure parameter can affect the price formation process.

\subsection{Two types of traders}
\label{part:twotypes}

To shed light on the EBS price clustering we start by noting that market participants can be divided into two groups: manual traders and automatic traders (computers algorithms). Then, we analyze the reaction of each of these groups to the decimalization. According to \citet{King2012} there is now a $50/50$ split (in orders volume) between algorithmic traders and manual traders with a keypad. We know from discussions with traders working at major banks that manual traders do not care about price improvement to that last decimal point if they are trying to trade in large sizes. Besides, they have been used to pip-pricing for many years and are not eager to adapt to the new system. On the contrary, automatic traders adapted quickly to the new tick size (just an algorithm adjustment) and take advantage of manual trading conservatism. They anticipate clustering tendencies and easily obtain priority by posting limit orders just above the best bid or just below the best ask\footnote{We can also add that of course traders are reluctant to post orders just behind clusters, it would imply losing price priority over the large depth available at integer prices.} More precisely, some traders try to genuinely take priority in order to deal at better price (the proportion of deals with last digit $1$ is above $10 \%$) while others are simply practicing flash trading\footnote{Flash traders send an order at the top of the book (new best price) followed by a cancellation to lure the other book observers. The goal is to make people believe that there is a bid (during the $250$ ms minimum quote life) at a certain level to trigger sales at this level. Those who give themselves a small margin to be sure of making the sale by showing a lower price to bid will automatically be executed with lower bids left in the book by the flash trader.}. This explains the $-1$ strong value in the limit orders placement distribution (figure \ref{figure:placement}) and the prominence of $1$ in the distribution of last digits in limit order prices (figure \ref{figure:limit_clust}). We now present two arguments which corroborate the analysis above.

\begin{figure}
\begin{center}
\includegraphics[width=0.5\textwidth,angle=-90]{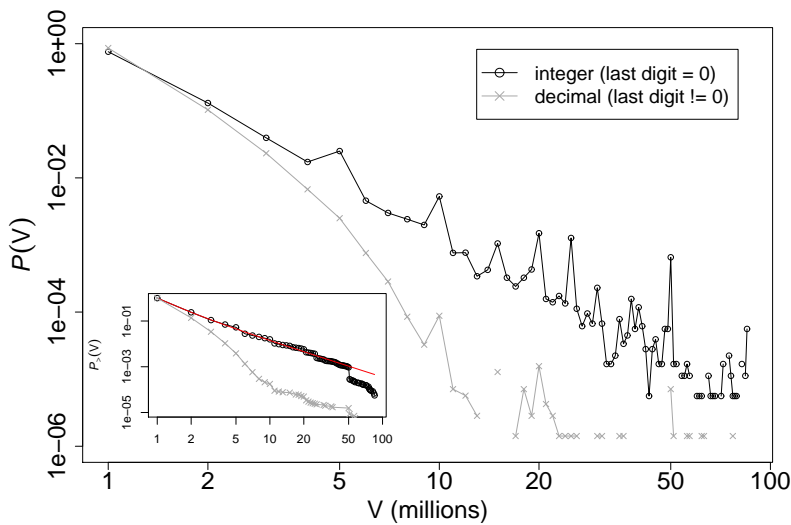}
\end{center}
\caption{Distribution of EUR/USD limit orders volume. The distribution depends on the order price last digit. Power-law for integer orders and exponential for decimal orders. The power-law exponents are stable through time: $2.6$ for EUR/USD and $2.7$ for USD/JPY. Inset: Cumulative distribution with power-law fit (discrete maximum likelihood estimator).}
\label{figure:twotypes}
\end{figure}

Firstly, some clustering effects are expected right after the decimalization, but they should decrease regularly as traders get used to the new tick. However, in the case of EBS the clustering is strong and stable. Therefore, part of the traders took into account the new situation whereas others did not and will not. Secondly, it is enlightening to look at the order volume depending on their price last digit. Figure \ref{figure:twotypes} plots in log-log scale the distribution of limit order volume in which the data were previously split into two samples: orders with integer price and orders with non-integer price. For integer orders the volume is broadly distributed. A discrete power-law maximum likelihood estimator returns exponents $2.6$ for EUR/USD and $2.7$ for USD/JPY (smaller values were reported for NASDAQ stocks, see \citep{Maslov2001}). Moreover, we observe peaks on big round volumes (5,10,15,20... millions) which is a trace of manual trading. In banks, large volume deals with customers are usually left to human dealers, therefore they accumulate large positions and they need to submit large orders to EBS to reduce their exposure quickly. On the other hand, for decimal orders the distribution is exponentially decreasing, typical of algorithmic trading. Indeed, automated market making systems are designed to avoid the accumulation of a large inventory and even if they have to liquidate a large position, they split it into small orders to limit their market impact.

\subsection{EUR/USD post-decimalization spread and clustering}
\label{part:spread_clust}

\begin{figure}
\begin{center}
\includegraphics[width=0.7\columnwidth]{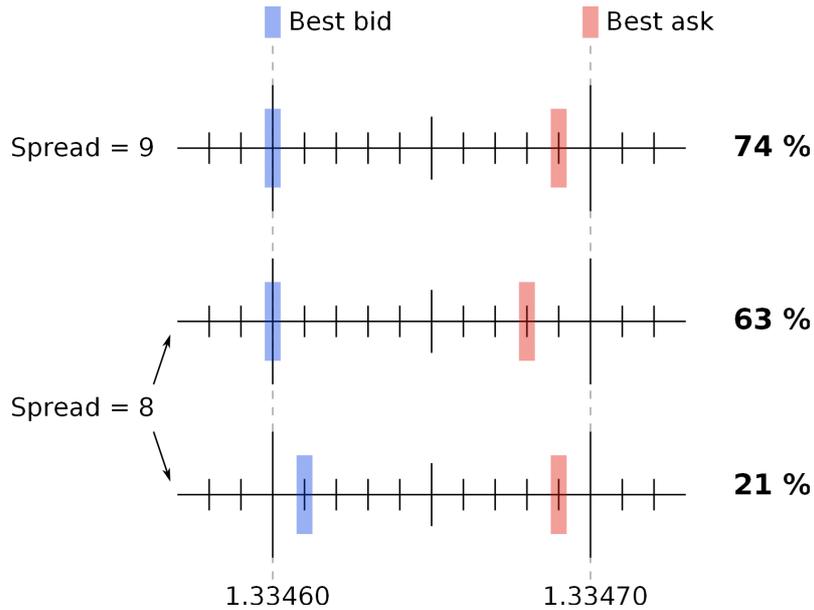}
\end{center}
\caption{Small spreads typology. Price clustering leads to a drastic limitation of spread configurations in terms of bid and ask last digit. We do not draw symmetric configurations (ask on integer instead of bid) but we count them. If the configurations were equally likely one would expect $20 \%$ per configuration for $9$ (odd spread : 5 possible configurations) and $16.67 \%$ per configuration for $8$ (even spread : 6 possible configurations), instead of the observed higher percentage.}
\label{figure:config1}
\end{figure}

The EUR/USD spread bimodality (section \ref{part:spread}) is intriguing and deserves further attention. In order to qualitatively understand the shape of the spread distribution, it is important to notice that the integer preference (or, equivalently, manual traders behavior) leads to a "natural" spread value of $10$, when the best bid and the best ask are on integers  (it corresponds to the pre-decimalization minimal spread). Then, the "step-ahead" (section \ref{part:twotypes}) strategy explains the first peak at $9$ and the large value at $8$. To confirm this hypothesis, we look at the quotes when the spread is $8$ or $9$. In theory, this can happen in a number of scenarios, depending on the last digits of bid and ask quotes. However, in our dataset we observe only one or two cases (see figure \ref{figure:config1}). For example, when the spread is equal to $9$, one of the best quotes of the configuration is an integer price for $74 \%$ of the time.

The second hump arises from the combination of clustering at integer values for one side (bid or ask) with the smaller clustering at half-integer values for the other side. Once again, some traders take price priority by posting limit orders just above or just below price barriers, which favors spreads slightly below $15$. This second peak in spread distribution is not as high as the first one because half-integer clustering is weaker than integer clustering and there is a natural tendency for the spread to revert to smaller values near $10$\footnote{Due to the existence of market making effects (also known as order book resiliency).}.

 \section{Conclusion}
In this paper we have reported the main outcome of a tick size change in the interdealer FX limit order book: an ubiquitous price clustering in both limit orders and actual trades. Interestingly, the uneven use of allowed price fractions can be explained by the interplay between manual traders, who stick to the old paradigm, and fast-adapting algorithmic traders. The most important consequence of this phenomenon is the appearance of support and resistance levels at integer and half-integer prices, which are not motivated by economic factors. We also note a strong distortion in the average shape of the book and in the spread distribution.

Our study led naturally to the question of what the optimal tick size for the EBS market is. On the one hand, the pip-pricing is praised by traditional dealing banks because it maintains a minimum level of profits and their traders are used to it. On the other hand, hedge-funds prefer finer grids because it reduces transaction costs, thus making it easier to use high-frequency strategies. Therefore, EBS market designers have to find a trade-off that satisfies both communities in order to maximize traded volume on their platform. From an empirical point of view, each tick size has advantages and drawbacks. When the old tick size was in place, the book exhibited high liquidity at each level and stable prices but the spread was almost always equal to one pip, suggesting that a tick size reduction was required. In contrast, the decimal pricing lowers the spread but generates a strong price clustering and allows the advent of high-frequency disruptive practices (e.g. flash trading). The USD/JPY situation is slightly better than the EUR/USD one: smaller spread and less pronounced price barriers. One possible explanation is that the  USD/JPY relative tick size\footnote{By relative tick size we mean $\delta = \frac{tick}{price}$. We find $\delta_{eurusd} \simeq 7\e{-6}$ and $\delta_{usdjpy} \simeq 1.2\e{-5}$, so $\frac{\delta_{usdjpy}}{\delta_{eurusd}} \simeq 1.7$.}  is bigger than the EUR/USD one. Our findings suggest that the optimal tick size lies somewhere between pip and decimal and show that taking into account traders biases is essential to design efficient trading structures. Remarkably, in September 2012, during the writing of this paper, EBS decided to change again the tick size and to go for half-pip pricing. It would be relevant to see if it managed to reduce price clustering while maintaining a small spread. We leave this question for future studies.

\section*{Acknowledgments}
We would like to thank Damien Challet,  Guillaume Pons, Aymen Jedidi and Ilija Zovko for helpful discussions and suggestions. We would also like to thank Joao da Gama Batista for his invaluable proofreading. All remaining mistakes are ours.

\newpage
\bibliographystyle{abbrvnat}
\bibliography{paper_one}

\end{document}